\documentclass[prl,superscriptaddress,floatfix,letterpaper,twocolumn]{revtex4}

\usepackage[utf8]{inputenc}

\usepackage{natbib}
\usepackage{amsmath}
\usepackage{amssymb}

\usepackage{graphicx}

\usepackage{verbatim}

\begin{document}

\title{Dispersive Photon Blockade in a Superconducting Circuit}

\author{A.J. Hoffman}
\author{S.J. Srinivasan}
\affiliation{Department of Electrical Engineering, Princeton University, Princeton, NJ 08544, USA}
\author{S. Schmidt}
\affiliation{Institute for Theoretical Physics, ETH Zurich, CH-8093 Zurich, Switzerland}
\author{L. Spietz}
\author{J. Aumentado}
\affiliation{National Institute of Standards and Technology, Boulder, CO 80305, USA}
\author{H.E. T\"{u}reci}
\author{A.A. Houck}
\affiliation{Department of Electrical Engineering, Princeton University, Princeton, NJ 08544, USA}

\date{April 14, 2010}

\begin{abstract}
Mediated photon-photon interactions are realized in a superconducting coplanar waveguide cavity coupled to a superconducting charge qubit.  These non-resonant interactions blockade the transmission of photons through the cavity.  This so-called dispersive photon blockade is characterized by measuring the total transmitted power while varying the energy spectrum of the photons incident on the cavity.  A staircase with four distinct steps is observed and can be understood in an analogy with electron transport and the Coulomb blockade in quantum dots.  This work differs from previous efforts in that the cavity-qubit excitations retain a photonic nature rather than a hybridization of qubit and photon and provides the needed tolerance to disorder for future condensed matter experiments.
\end{abstract}


\maketitle

For decades, light has served as a useful tool in condensed matter physics, yet rarely has light itself been studied in this same framework ~\cite{Jaksch1998, Stamper-Kurn1998}.  The reason that light has been relegated to a tool of condensed matter physics, rather than a subject, is that photons do not interact, and even mediated interactions are typically weak.  However, strong nonlinearities on the single photon level arising from the interaction between photons and an atom or qubit in a cavity have been realized~\cite{Birnbaum2005, Bishop2009, Fink2008, Faraon2008}.  Recently, several proposals founded on these strong mediated interactions have been set forth to study strongly correlated macroscopic systems with interacting photons or polaritons in arrays of cavities coupled to atoms or qubits~\cite{Greentree2006, Hartmann2006, Angelakis2007, Rossini2007, Hartmann2008, Na2008, Schmidt2009, Tomadin2010}.  A single element of these lattices, a cavity with mediated photon-photon interactions, is studied here.  For sufficiently strong interactions, photon transmission through the cavity can be significantly modified.

Resonant photon blockade, motivated by Coulomb blockade in condensed matter physics~\cite{Fulton1987, Kastner1992}, was first observed in an optical cavity coupled to a single trapped atom~\cite{Rebic1999, Tian1992, Imamoglu1997}.  As the atom is brought into resonance with the cavity, the resulting energy spectrum is nonlinear and photon transmission becomes sub-poissonian and anti-bunched~\cite{Birnbaum2005}.  The resonant blockade was also later demonstrated in a photonic crystal cavity coupled to a quantum dot~\cite{Faraon2008}.  Recently, strong resonant nonlinearities have also been observed spectroscopically in atoms~\cite{Schuster2008} and in circuit quantum electrodynamics (cQED), in which a superconducting qubit is coupled to a microwave transmission line cavity~\cite{Wallraff2004}.  Here, the $\sqrt{n}$ spacing of the Jaynes-Cumming ladder was observed and used to generate nonclassical states of light~\cite{Schuster2008, Fink2008, Hofheinz2008, Bishop2009}.  In all of this previous work, strong photon-photon interactions were observed when the qubit and cavity were resonant, resulting in highly hybridized excitations between the qubit and photon.

Here, we demonstrate a dispersive photon blockade, where an off-resonant superconducting qubit coupled to a cavity provides sufficient non-linearity to result in quantized transmission of photons through the cavity.  In this dispersive regime, the excitations of the cavity-qubit system are photonic and the resulting nonlinearity is a result of mediated photon-photon interactions.  A measurement of the total transmitted power through the cavity while increasing the incident photon bandwidth results in a staircase, where each step indicates that an additional photon can be present in the cavity.  The dispersive interaction can be thought of as a qubit-mediated photon-photon interaction, and the strongly nonlinear transmission can be understood in analogy to transport in confined electronic structures~\cite{Kastner1992}.  This dispersive photon-photon interaction is a key step towards the study of condensed matter physics with photons.

Our system consists of a superconducting qubit off-resonantly coupled to a microwave cavity.  The cavity is driven incoherently by modulating a coherent tone with frequency $\omega_{\rm L}$ with band-limited Gaussian white noise that has a cutoff frequency $\sigma$. In the dispersive approximation \cite{Boissonneault2009}, where the qubit remains in its ground-state, the dynamics of the system is described well by the effective Hamiltonian $H=H_0+H_{\rm dr}$.  Here, $H_0=\delta a^\dagger a + U (a^\dagger a)^2$, is the system Hamiltonian with detuning $\delta=\omega_{\rm eff}-\omega_{\rm L}$ and photon interaction strength ,$U$, which is obtained from exact diagonalization of a multi-level system with generalized Jaynes-Cummings interaction. \footnote{In the numerics leading to Fig. 2 we take into account the slight dependence of the effective cavity frequency $\omega_{\rm eff}$ and $U$ on the number of photons present in the cavity~\cite{Boissonneault2010}}.  The second term, $H_{dr} = \eta(t)(a+a^\dagger)$, introduces the band-limited Gaussian white noise drive described by $\overline{\eta(t)}=0,\,\, \overline{\eta(t)\eta(t')}=f\, C_\sigma(t-t')$ with variance $f$ and the correlator $C_\sigma(\tau)=\sin(\sigma\tau)/\tau$.  The photon-photon interaction modifies the energy required to add a photon to the cavity depending upon the number of photons occupying the cavity.  Although this is a higher order effect, the interaction energy, or Kerr energy, can be made larger than the cavity linewidth, $\kappa$, and the transport of photons through the cavity is blockaded.

\begin{figure}
	\includegraphics[width=0.4\textwidth,clip]{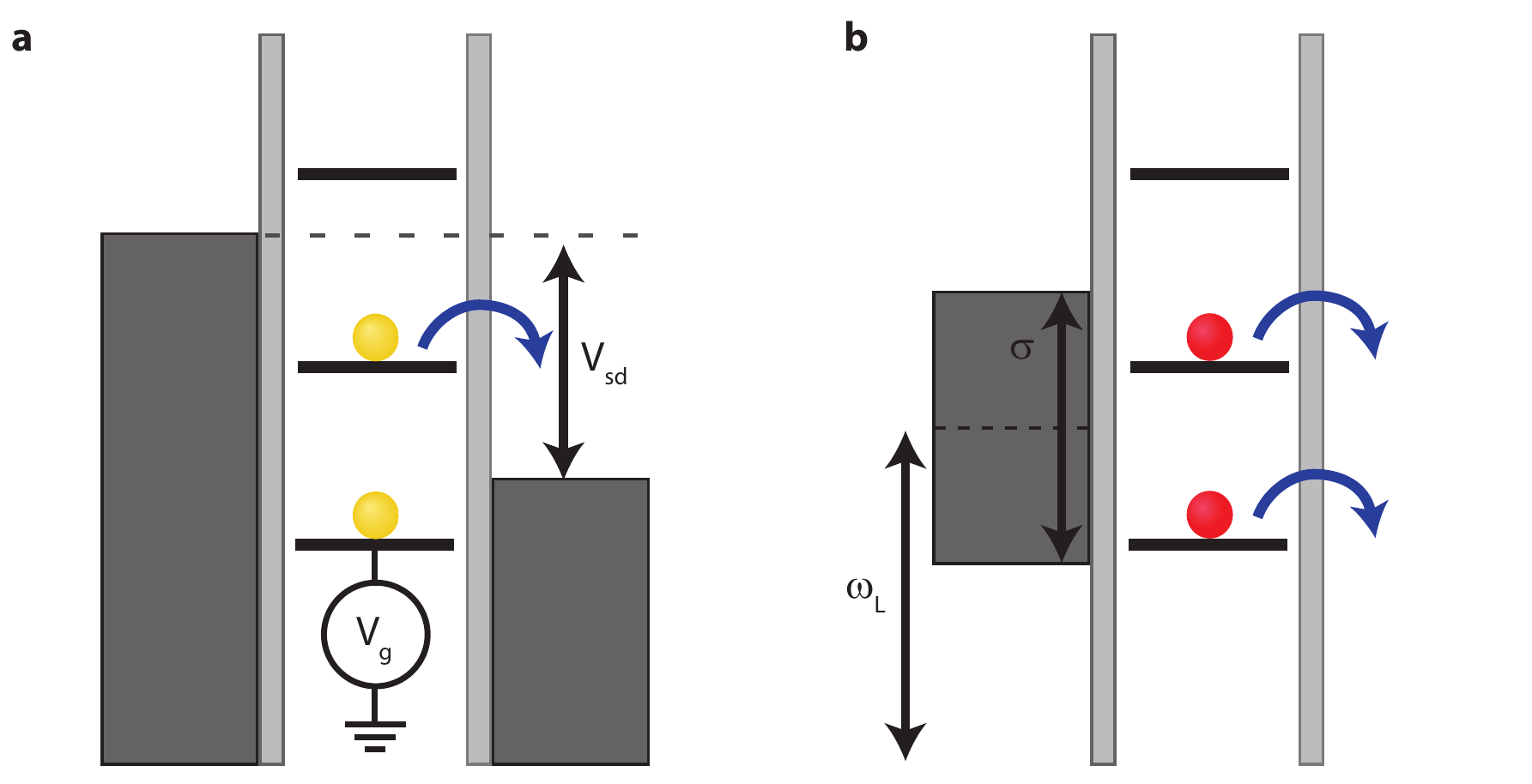}
	\caption{\label{figure1} Energy diagrams for Coulomb and photon blockades, with grey rectangles representing the distribution of particle energies in the leads. (a) Schematic for electronic transport through a quantum dot.  The gate voltage, $V_g$  controls the energy offset between states in the dot and electrons in the leads, while the source-drain voltage, $V_{\textit{sd}}$, sets the range of electron energies that participate in transport.  The Coulomb blockade can be overcome by increasing $V_{\textit{sd}}$.  (b) Schematic for transmission through a dispersively coupled cavity-qubit system in the photon blockade regime.  The center frequency, $\omega_{\textit{L}}$, can be changed to align the energy of incident photons with states of the cavity, much like $V_g$ in a quantum dot.  The bandwidth of the incident photons, $\Delta \omega$, controls the energy range of the photons in the system and can be increased to overcome photon blockade.}
\end{figure}

This system, in which transport of photons can be blockaded due to interactions, bears strong resemblance to a confined electronic quantum dot, a canonical condensed matter system in which transport of electrons can be similarly blockaded due to Coulomb interactions~\cite{Fulton1987, Kastner1992}.  Both systems consist of a set of discrete energy levels with large interaction energy that depends on $n^2$, where $n$ represents the number of particles.  The Kerr energy for photon blockade is equivalent to the charging energy in a quantum dot, and both represent the additional energy required to add a particle to the confined structure due to interactions. With this similarity in mind, the physics of the photon blockaded system can be probed in a way akin to that in a Coulomb blockaded system.  The response of the blockaded cavity to an applied microwave stimulus mimics that of a quantum dot to applied voltages.  A range of incident photon energies, similar to a Fermi sea of electrons, is incident on the resonator.  Because only microwaves in a narrow energy range are applied, a measurement of total transmitted power is a good proxy for the total photon number current through the device, a quantity similar to the electron current through a quantum dot.

Figure \ref{figure1} shows diagrams of the energy configurations of the photon and electron systems.  Transport through the cavity is only possible when the frequency of applied radiation matches the frequency of the empty states in the cavity; thus the excitation frequency plays the role of the gate voltage typically used in electron transport experiments to align states in the leads with those in the dot.  A broader range of photon energies can be used to overcome photon blockade, just as an applied source-drain voltage can overcome Coulomb blockade or allow current transport through multiple excited electronic states in a quantum dot.  Therefore, by measuring the total transmitted power while varying the center frequency and bandwidth of an incoherent excitation, it is possible to measure transmission maps similar to Coulomb blockade diamonds.

\begin{figure}
	\includegraphics[width=0.43\textwidth,clip]{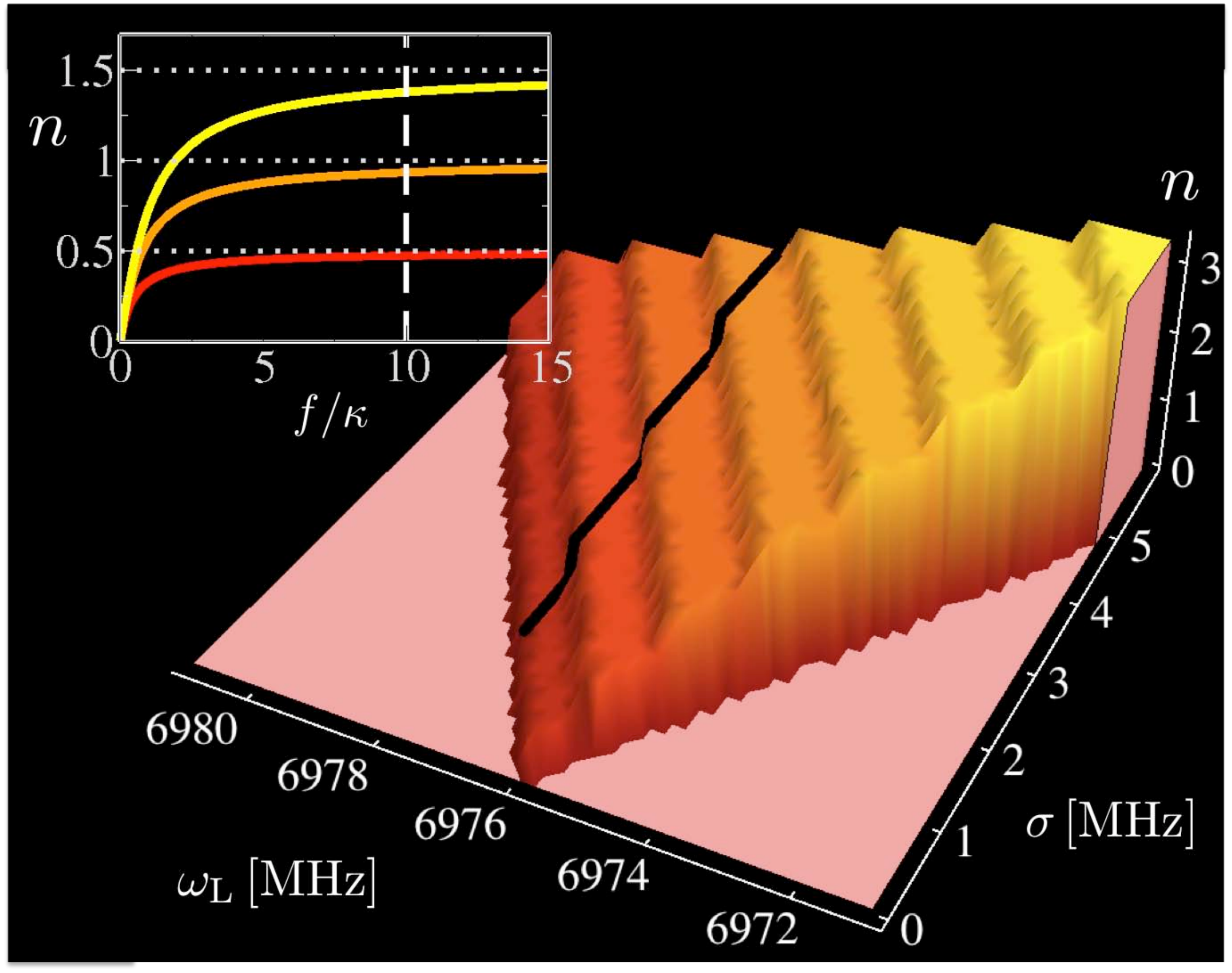}
	\caption{\label{figure2}Calculated number of photons, $n$, versus incident photon bandwidth, $\sigma$, and center frequency, $\omega_{\rm L}$, the system parameters quoted in the text. The black line shows a typical photon staircase for $\omega_{\rm L}=6977\, \mathrm{MHz}$. The inset shows the number of photons as a function of effective drive $f/\kappa$ on the first three plateaus of the staircase function, i.e., for $\omega_{\rm L}=6977 {\rm MHz}$ and $\sigma=1.5\, \mathrm{MHz}$ (red), $\sigma=2.5\, \mathrm{MHz}$ (orange) and $\sigma=4\, \mathrm{MHz}$ (yellow). Here, the white dashed line corresponds to the value $f/\kappa = 10$ used in the main graph.}
\end{figure}

There are important differences between photon and electron transport. Unlike electrons and a true femionic reservoir in the leads, the incident photons are bosonic waveguide modes that have a distribution function determined by the band-limited white noise. It is the absence of low energy incident photons, rather than an absence of unoccupied states in the output (drain), that prevents transport when all of the incident photon energies are too high; given the bosonic nature of photons, there is always an available state for photons exiting the cavity. Furthermore, in blockaded quantum transport, one typically has the condition that the charging energy is much larger than the mean-level spacing of single particle levels, each of which can in turn accommodate two electrons. In the photon blockade regime studied here, the opposite situation exists, where the higher energy levels are cavity harmonics with spacings much larger than $U$.  Hence it is one single-particle level that can accommodate any number of photons that is active in the regime explored here.

Figure \ref{figure2} shows the calculated number of photons in the cavity for a dispersive photon blockade plotted in a way reminiscent of charge stability diagrams in quantum transport. The results are obtained from the steady state density matrix $\rho_{\rm s}$ of the system. To leading order in the drive (Born approximation) $\rho_s$ is determined by the equation
\begin{eqnarray}
\label{rhoeqn}
0=-i[H_0,\rho_{\rm s}]+f(\mathcal{L}^{\downarrow}_\sigma[\rho_{\rm s}]+\mathcal{L}^{\uparrow}_\sigma[\rho_{\rm s}]) + \kappa\mathcal{L}^{\downarrow}_\infty[\rho_{\rm s}]
\end{eqnarray}
with generalized Lindblad operators
\begin{eqnarray}
\mathcal{L}^{\uparrow}_\sigma[\rho_{\rm s}] = a \rho_{\rm s} a_\sigma^\dagger + a_\sigma \rho_{\rm s} a^\dagger - a^\dagger a_\sigma\rho_{\rm s} - \rho_{\rm s} a^\dagger_\sigma a
\end{eqnarray}
and
\begin{eqnarray}
\mathcal{L}^{\downarrow}_\sigma[\rho_{\rm s}] = a_\sigma^\dagger \rho_{\rm s} a + a^\dagger \rho_{\rm s} a_\sigma  - a a_\sigma^\dagger\rho_{\rm s} - \rho_{\rm s} a_\sigma a^\dagger
\end{eqnarray}
and the ladder operators
$a_\sigma=\sum_{n} \sqrt{n}\, \mathcal{S}_\sigma (\Delta_{n}) |n\rangle \langle n-1|$,
where $|n\rangle$ are photon Fock states with occupation number $n$, $\Delta_{n}=\epsilon_n-\epsilon_{n-1}$ is given in terms of the eigenvalues of $H_0$ as $\epsilon_n=\delta \cdot n +U n(n-1)$ and
$\mathcal{S}_\sigma(\Delta_n)$ denotes the noise spectral function defined by
$\mathcal{S}_\sigma (\omega)=\int^\infty_{-\infty} C_\sigma(\tau) e^{i\omega\tau}d\tau$.
The first term on the r.h.s in (\ref{rhoeqn}) describes the coherent evolution under the system Hamiltonian $H_0$, the second term introduces an energy dependent pump and decay due to the incoherent drive with rate $f$, and the last term describes photon loss.
Note, that for a white noise bath with $\sigma\rightarrow\infty$, we recover the usual boson operator $a_\infty =a$ and Lindblad operator
$\mathcal{L}^{\downarrow}_\infty =2 a^\dagger \rho_{\rm s} a  - a a^\dagger\rho_{\rm s} - \rho_{\rm s} a a^\dagger$.
In order to calculate the number of photons  $n={\rm Tr} [a^\dagger a] \rho_{\rm s}$ in the cavity we solve Eq.~(\ref{rhoeqn}) for $\rho_{\rm s}$ in the eigenbasis of $H_0$ numerically.

This physics of photon blockade studied above is experimentally probed using a cQED sample~\cite{Blais2004, Wallraff2004, Frunzio2005}.  The superconducting transmission line cavity has a $\lambda/2$ resonance at $\omega_r/2\pi = 6.91\,\mathrm{GHz}$ that overlaps with the ultra-low noise bandwidth range of a following microwave SQUID amplifier and a narrow cavity linewidth, $\kappa/2 \pi \approx 100\,\mathrm{kHz} $.A superconducting transmon qubit~\cite{Houck2009} provides the nonlinearity for the photon-photon interaction, with a qubit-photon coupling strength, $g/2\pi=260\,\mathrm{MHz} $, measured using the peak separation in a qubit number splitting experiment~\cite{Schuster2006}, which gives rise to the photon-photon interaction strength, $U \sim 1\, \mathrm{MHz}$.

\begin{figure}
	\includegraphics[width=0.41\textwidth,clip]{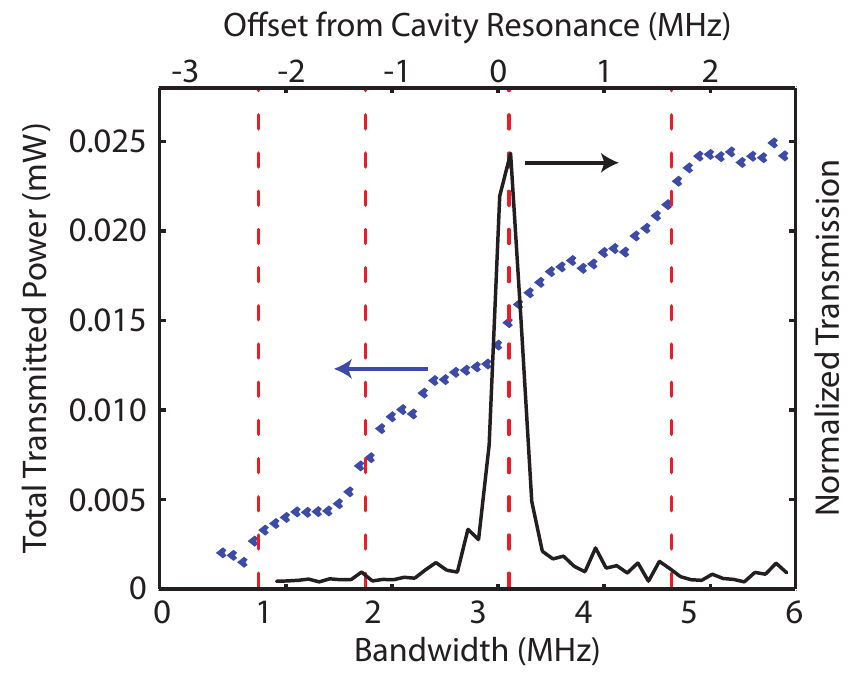}
	\caption{\label{figure4}Incoherent and coherent cavity transmission. The blue dots show the measured power transmitted through the cavity-qubit system in a 8 MHz measurement window for different incident photon bandwidths. The data was over-sampled in bandwidth and the noise was reduced by using a five-point smoothing function. Four steps in the transmission are evident and correspond to the transmission of an additional photon.  The vertical dashed lines indicate step locations based on eigenstates obtained by diagonalizing the Jaynes-Cummings Hamiltonian.  Rounding of the steps in the staircase is consistent with measured $Q$ of the cavity.  The measurement error is about 5\%.  The solid black line shows the normalized transmission spectrum measured using a coherent tone at few-photon power levels.}
\end{figure}

A photon staircase, the signature of dispersive photon blockade, is observed by measuring the total transmitted power in a large fixed bandwidth (8 MHz) while varying the width of the incident photon spectrum.  In our experiment, incident photons are supplied by an incoherent, broadband RF tone generated by modulating a coherent tone with white Gaussian noise in a given bandwidth; the spectrum of available photon energies is controlled by varying the bandwidth of the noise, and thus the width of the incident photon spectrum.  When the device is blockaded, transmission is on the few-photon level and occupies only a small fraction of the total measured bandwidth, resulting in exceptionally low signal compared with amplifier noise.  To increase the sensitivity of our measurement,  this experiment employed an ultra-low noise SQUID amplifier designed to operate in a 10 MHz bandwidth that matches the frequency of the cavity when accounting for the shift due to the presence of a qubit~\cite{Spietz2008, Spietz2009}.  This device has a system noise temperature of ~1K and provides the dominant source of noise in the measurement.  In addition, to minimize error in the measurement due to drift in the gain and noise floor of the amplifiers, the input signal is amplitude modulated and the difference between the on and off state is calculated.  A heavily averaged series of measurements with an applied spectral power density of -147 dBm/MHz are taken for bandwidths ranging from 2 - 12 MHz.

The measured data, demonstrating the signature photon blockade staircase, are presented in Figure \ref{figure4}(a).  When the incident bandwidth is sufficiently large, incident photons have enough energy to overcome the interaction energy, and simultaneous transport of individual photons at multiple energies becomes possible.  Four steps in the transmitted power are clearly resolvable as the bandwidth is increased, each corresponding to transport of an additional photon and occurring at bandwidths that match theoretical predictions.   The vertical dashed lines show the predicted step locations due to the nonlinearity found using the Jaynes-Cummings Hamiltonian and independently recovered device parameters, $g/2\pi = 260\,\mathrm{MHz}$, $Q = 70000$, and $\Delta/2\pi = 0.96\,\mathrm{GHz}$.  There is excellent agreement between the numerical calculations and the measured results.  The rounding in the steps is due to the spectral width of the cavity and can be improved by increasing the $Q$.
Several control experiments were also performed to verify the blockade.  Transmission measurements using a weak coherent tone revealed a narrow cavity with no spurious resonances.  An identical broadband measurement on a sample without a qubit did not reveal a staircase.  
Recent advances in circuit QED have enabled measurements of the time correlation between transmitted photons~\cite{Silva2010}.  Such work could possibly be extended to study the broadband excitations used here and is an important subject for future work.

The dispersive photon blockade demonstrated in this work provides an important step towards the study of condensed matter physics with qubit-mediated photon-photon interactions.  The observed steps in the total transmitted power with increasing incident photon bandwidth demonstrate that a qubit coupled to a high $Q$ cavity can blockade transmission.  This dispersive photon blockade is the result of mediated photon-photon interactions, the magnitude and sign of which can be controlled via the qubit energy and be made larger than the cavity linewidth.  While here we study a rather simple system consisting of strong interactions in one cavity, it should be possible to array many such devices in a one- or two-dimensional lattice.
The lithographic nature of these devices allows creation of many different lattice geometries, and the sign and magnitude of the interactions can be dynamically tuned by changing qubit energies.  Moreover, the dispersive interaction demonstrated here, implies a strong photon-photon interaction over a wide range of frequencies, thus rendering complicated devices less susceptible to disorder.

\begin{acknowledgments}
	The authors acknowledge fruitful discussions with Jens Koch, David Schuster, G. Blatter, and A. Imamoglu.  The Princeton work was supported by the NSF CAREER Award (Grant No. DMR-0953475), the Sloan Foundation, the Packard Foundation, the NSF Funded Princeton Center for Complex Materials (Grant No. DMR-0819860).  Work at NIST was funded by the NSA. S.S. thanks the Princeton Center for Theoretical Science (PCTS) for its hospitality and acknowledges support from the Swiss NSF under Grant No. IZK0Z2\_134286.
\end{acknowledgments}

\bibliography{QED}

\end{document}